\documentstyle[12pt]{article}
\begin{document}
\date{\today}
%
%Naturais
\def\N{\rm l\!N\,}
%Inteiros
\def\Z{\angle \!\!\!{\rm Z}}
%Reais
\def\R{\rm l\!R\,}
%Racionais
\def\Q{\rm l\!\!\!Q\,}
%Complexos
\def\C{\rm l\!\!\!C\,}
%Quaterni\~oes
\def\H{\rm l\!H\,}
%Octoni\ões
\def\O{\rm l\!\!\!O\,}
%Unidade matricial
\def\uma{\rm 1\!\!\hskip 1 pt l}
%Comprimento de onda de Compton
\def\Compton{\overline{\phantom{\overline{.\!.}}}\!\!\!\lambda}
%
%
%%%%%%%%%%%% Casimir effect at finite temperature of
%                        charged scalar field in an external magnetic field %%%%%%%%%%%%
%
%
\title{Casimir effect at finite temperature of charged scalar field in
an external magnetic field}
\author{M. V. Cougo-Pinto\thanks{e-mail: marcus@if.ufrj.br},
C. Farina\thanks{e-mail: farina@if.ufrj.br}, M. R.
Negr\~ao\thanks{e-mail: guida@if.ufrj.br} and A.
Tort\thanks{e-mail: tort@if.ufrj.br}\\
\\Instituto de F\'\i sica, Universidade Federal do Rio de
Janeiro\\ CP 68528, Rio de Janeiro, RJ 21945-970, Brazil\\
\\}
\maketitle
\begin{abstract}
The Casimir effect for Dirac as well as for scalar charged
particles is influenced by external magnetic fields. It is also
influenced by finite temperature. Here we consider the Casimir
effect for a charged scalar field under the combined influence of
an external magnetic field and finite temperature. The free energy
for such a system is computed using Schwinger's method for the
calculation of the effective action in the imaginary time
formalism. We consider both the limits of strong and weak magnetic
field in which we compute the Casimir free energy and pressure.
\end{abstract}
%%%%%%%%%%%%%%%%%%%%%%%%%%%%%%%%%%%%%%%%%%%%%%%%%%%%%%%%%%%%
%
The Casimir effect can be generally defined as the effect of
non-trivial space topology on the vacuum fluctuations of
relativistic quantum fields
\cite{MostepanenkoTrunovB,PlunienMullerGreinerR}. The corresponding
change in the vacuum fluctuations appears as a shift in the vacuum
energy and a resulting vacuum pressure. In the original Casimir
effect \cite{Casimir48} two parallel closely spaced conducting
plates confine the electromagnetic field vacuum in the region
between the plates. We may consider the plates as squares of side
$\ell$ separated by a distance $a$; the close spacing is
implemented by the condition $a\ll\ell$. A shift in the zero point
energy, known as the Casimir energy, is produced in passing from
the trivial space topology of $\R^3$ to the topology of
$\R^2\times[0,a]$. As a consequence, the plates are attracted
towards each other, albeit being uncharged. This force of
attraction has been measured by Sparnaay \cite{Sparnaay58} and
recently with high precision by Lamoreaux \cite{Lamoreaux97} and by
Mohideen and Roy \cite{MohideenRoy98}. The slab of vacuum between
the plates can be seen as a system with large volume $a\ell^2$,
energy given by the Casimir energy and pressure given by minus the
derivative of the Casimir energy with respect to the spacing $a$.
It is then natural to look for the thermodynamical properties of
such a system by considering its behavior at finite temperature
\cite{Fierz60}. The system at temperature $1/\beta$ can be
described by its partition function $Z(\beta)$ in terms of which we
obtain the free energy $F(\beta)=-\beta^{-1}\log{Z(\beta)}$. In the
limit of zero temperature $F(\infty)$ gives the usual Casimir
energy. The Casimir effect has been computed for fields other than
the electromagnetic and boundary conditions different from the one
implemented by conducting plates
\cite{MostepanenkoTrunovB,PlunienMullerGreinerR}. In those cases
also it is important to consider the finite temperature effects.
Here we are interested in the case of a charged scalar field at
finite temperature under the combined influence of confinement and
an external magnetic field. The confinement is implemented by two
impermeable parallel plates, as in the usual electromagnetic
Casimir effect described above, and the external magnetic field
${\bf B}$ is constant and perpendicular to the plates. The
influence of the external magnetic field on the Casimir effect has
already been computed \cite{Bsc1Caxa97} and here we consider the
combined effect of external magnetic field and finite temperature
on the Casimir effect. In our calculation we use Schwinger's
formula for the effective action \cite{Schwinger51}, which can be
used to calculate the Casimir energy \cite{Schwinger92}. It can be
used also to compute the partition function $Z(\beta)$
\cite{CPFT96} in the imaginary time formalism \cite{DolanJackiw74}
for finite temperature $1/\beta$:
\begin{equation}\label{Z}
\log{Z(\beta)}={1\over 2}\int_{s_o}^\infty
{ds\over s}\;Tr\,e^{-isH}\; ,
\end{equation}
where $s_o$ is a cutoff in the proper-time $s$, $Tr$ means the
total trace and $H$ is the proper-time Hamiltonian in which the
frequencies have been discretized to the values $i2\pi n/\beta$
($n\in\Z$). For the charged scalar field we have
$H=(-i\partial-eA)^2+m^2$, where $e$ and $m$ are the charge and
mass of the field. We have for the trace in (\ref{Z}):
\begin{eqnarray}\label{Tr}
Tr\,e^{-isH}= 2e^{-ism^{2}}\sum_{n'=1}^{\infty}{eB\ell^2\over
2\pi}e^{-iseB(2n'+1)}
\sum_{n_{1}=1}^{\infty}e^{-is(\pi {n_{1}}/a)^2}
\sum_{n_{2}=-\infty}^{\infty}
e^{-is(2\pi n_{2}/\beta)^2} ,
\label{Tr}
\end{eqnarray}
where the first factor 2 is due to the charge multiplicity, the
first sum is on the Landau levels with the corresponding degeneracy
factor, the second sum is on the eigenvalues stemming from the
condition of confinement between the plates and the third sum is
due to the finite temperature condition. The sum on the Landau
levels is straightforward and the other two sums can be modified by
using Poisson summation formula \cite{Poisson1823}:
\begin{eqnarray}
\sum_{n=-\infty}^{\infty} e^{-n^2\pi\tau}={1\over\sqrt{\tau}}
\sum_{n=-\infty}^{\infty} e^{-n^2\pi/\tau} .
\label{Poisson}
\end{eqnarray}
The trace is then given by:
\begin{eqnarray}
Tr\,e^{-isH}=
-\beta \frac{a\ell^2}{4{\pi}^{2}} \frac{e^{-ism^2}}{s^{2}}
[1 + iseB{\cal M}(iseB)]
\biggl [ \sum_{n_{1}=1}^\infty {e^{i(an_{1})^2/s}} -
\frac{\sqrt{i\pi s}}{2a} + \frac{1}{2} \biggl ]
\biggl [ 1 +2\sum_{n_{2}=1}^\infty e^{i(\beta n_{2})^2/4s}
\biggl ],
\label{Trfinal}
\end{eqnarray}
where we have used the function ${\cal M}$ defined by ${\cal
M}(\xi)=cosech{\xi}-\xi^{-1}$. By using this expression for the
trace in (\ref{Z}) and substituing the result into
$F(a,B,\beta)=-\beta^{-1}\log{Z(\beta)}$ we obtain:
%
%\hfill\eject
%
\begin{eqnarray}\label{FreeE}
F(a,B,\beta)&=&-a\ell^2
\left\{
{1\over 16\pi^2}\int_{s_o}^\infty {ds\over s^3}
e^{-ism^2}[1+iseB{\cal M}(iseB)]+\right.\nonumber\\ &+&\left.
2\sum_{n_{2}=1}^\infty {1\over 16\pi^2}
\int_{s_o}^\infty {ds\over s^3} e^{-ism^2+i(\beta n_{2}/2)^2/s}
[1+iseB{\cal M}(iseB)] \right\}+\nonumber\\ &+&\ell^2
\left\{ {1\over 16(i\pi)^{3/2}}\int_{s_o}^\infty {ds\over s^{5/2}}
e^{-ism^2}[1+iseB{\cal M}(iseB)]+\right.\nonumber\\ &+&\left.
2\sum_{n_{2}=1}^\infty {1\over 16(i\pi)^{3/2}}
\int_{s_o}^\infty {ds\over s^{5/2}} e^{-ism^2+i(\beta n_{2}/2)^2/s}
[1+iseB{\cal M}(iseB)] \right\}+\nonumber\\
&+&\sum_{n_{1}=1}^\infty \left\{ {a\ell^2\over
8\pi^{2}}\int_{s_o}^\infty {ds\over s^{3}}
e^{-ism^2+i(an_{1})^2/s}[1+iseB{\cal M}(iseB)]+\right.\nonumber\\
&+&\left. 2\sum_{n_{2}=1}^\infty {a\ell^2\over 8\pi^{2}}
\int_{s_o}^\infty {ds\over s^{3}} e^{-ism^2+i[(an_{1})^2+(\beta
{n_{2}}/2)^2]/s}[1+iseB{\cal M}(iseB)]
\right\}
\end{eqnarray}
The term proportional to the volume $a\ell^2$ comes from a uniform
energy density throughout space. This density does not depend on
the position of the plates and does not contribute to the Casimir
free energy. The term proportional to the area $\ell^2$ of each
plate is independent of the position of the plates and does not
contribute also to the Casimir free energy. Those two terms are
plagued by infinities when the cutoff $s_o$ is eliminated and
require the standard renormalizations in the energy $F$
\cite{Schwinger92} and also in the charge $e$ and field $B$
\cite{H-E,Schwinger51}. Those renormalizations do not affect the
form of other terms in the free energy (\ref{FreeE}) and we may
consider the following expression as the properly renormalized
Casimir free energy:
\begin{eqnarray}\label{CasiFreeE}
F_C(a,B,\beta)&=& {a\ell^2\over 8\pi^{2}}\sum_{n_{1}=1}^\infty
\left\{
\int_{s_o}^\infty {ds\over s^{3}}
e^{-ism^2+i(an_{1})^2/s}[1+iseB{\cal M}(iseB)]+\right. \nonumber\\
&+&\left. 2\sum_{n_{2}=1}^\infty
\int_{s_o}^\infty {ds\over s^{3}} e^{-ism^2+i[(an_{1})^2+(\beta
n_{2}/2a)^2]/s} [1+iseB{\cal M}(iseB)]
\right\}\; ,
\end{eqnarray}
In this expression the contribution which does not depend on the
external magnetic field can be integrated out (cf. formula {\bf
3.471},9 in \cite{Grad}) and (\ref{CasiFreeE}) can be rewriten as:
\begin{eqnarray}\label{CasiFreeE2}
{F_C(a,B,\beta)\over a\ell^2}&=&
-{(am)^2\over 4\pi^2 a^4}\sum_{n_{1}=1}^\infty {1\over
n_{1}^2}K_2(2amn_{1})+\nonumber\\ &+&{2(am)^4\over \pi^2
a^4}\sum_{n_{1},n_{2}=1}^\infty \frac{1}{(2amn_{1})^2+(\beta
mn_{2})^2} K_2\left(\sqrt{(2amn_{1})^2+(\beta
mn_{2})^2}\right)+\nonumber\\ &+&{1\over
8\pi^{2}}\sum_{n_{1}=1}^\infty
\left\{
\int_{s_o}^\infty {ds\over s^{3}}
e^{-ism^2+i(an_{1})^2/s} iseB{\cal M}(iseB)+\right. \nonumber\\
&+&\left. 2\sum_{n_{2}=1}^\infty
\int_{s_o}^\infty {ds\over s^{3}} e^{-ism^2+i[(an_{1})^2+
(\beta n_{2}/2)^2]/s} iseB{\cal M}(iseB)
\right\}
\end{eqnarray}
The external magnetic field appears in (\ref{CasiFreeE}) through
the function ${\cal M}$ and it has the effect of decreasing the
Casimir free energy. The effect is more pronounced in the strong
field regime, which we consider now. Let us notice that the
integrals in (\ref{CasiFreeE}) are dominated by the bell-shaped
exponential function, which has its maximum at $\sqrt{n^2+(\beta
n_{2}/2a)^2}/am$ (obviously with $\beta=0$ in the case of the first
integral). Therefore, the strong magnetic field regime is defined
by $eBa^2\gg am$ or, if we prefer, by $B\gg (\phi_o/
a^2)(a/\Compton)$, where $\phi_o$ is the fundamental flux $1/e$ and
$\Compton$ is the Compton wavelength $1/m$. In this strong field
regime we can substitute $1+\xi {\cal M}(\xi)$ by $2\xi e^{-\xi}$
and (\ref{CasiFreeE}) can be approximated by (cf. formula {\bf
3.471},9 in \cite{Grad}):
\begin{eqnarray}\label{CasiFreeEatB>>}
{F_C(a,B,\beta)\over a\ell^2}&=&
-{eB\over 2\pi^2 a^2}\sum_{n=1}^\infty {\sqrt{(am)^2+eBa^2}\over n_{1}}
K_1\left(2n_{1}\sqrt{(am)^2+eBa^2}n_{1}\right) +\nonumber\\
&-&{2eB(am)^2\over \pi^2 a^4}{\sqrt{(am)^2+eBa^2}
\sum_{n_{1},n_{2}=1}^\infty
\frac{K_1\left(\sqrt{(m^2+eB)[(2an_{1})^2+(\beta
n_{2})^2}\right)}{\sqrt{(2amn_{1})^2+(\beta mn_{2})^2}}}
\end{eqnarray}
By using this expression in the formula for the pressure,
$p(a,B,\beta)=\ell^{-2}\partial F_C(a,B,\beta)/\partial a$, we
obtain finally (cf. formula {\bf 8.472},3 in \cite{Grad}):
\begin{eqnarray}\label{CasipatB>>}
p_C(a,B,\beta)&=& {eB\over \pi^2}\sum_{n_{1}=1}^\infty (m^2+eB)
\biggl \{ K_0(2a\sqrt{m^2+eB}n_{1}) -
{K_1(2a\sqrt{m^2+eB}n_{1})
\over 2a\sqrt{m^2+eB}n_{1}} + \nonumber\\
&+& 2\sum_{n_{2}=1}^\infty
\biggl [ \frac{K_{1}(\sqrt{(m^2 + eB)[(2an_{1})^2 + (\beta
n_{2})^2]})}{\sqrt{(m^2 + eB)[(2an_{1})^2 + (\beta n_{2})^2]}
}+\nonumber\\ &-& 4an_{1}^{2} \frac{K_0( \sqrt{(m^2 +
eB)[(2an_{1})^2 + (\beta n_{2})^2]})}{(2an_{1})^{2}+(\beta
n_{2})^{2}}+
\nonumber\\
&-& 2(2an_{1})^{2}\frac{K_1(\sqrt{(m^2 + eB)[(2an_{1})^2 +
(\beta n_{2})^2]})}{\sqrt{m^2+eB}[(2an_{1})^{2}+(\beta
n_{2})]^{3/2}}
\biggl ] \biggl \}
\end{eqnarray}
In this expression the first sum gives the contribution to the
pressure from the strong magnetic field and the double sum gives
the combined effect form the strong field and the finite
temperature.

When we consider the weak field regime, $eBa^2 << am$ and we
substitute $1+
\xi {\cal M}(\xi)$ by $1-\frac{{\xi}^{2}}{6}\frac{7{\xi}^{4}}{360}$ and
(\ref{CasiFreeE}) can be approximated by:
\begin{eqnarray}
\frac{F_C(a,B,\beta)}{a\ell^{2}} &=& -\frac{m^{2}}{4\pi^{2}a^2}
\sum_{n_{1}=1}^{\infty} \frac{1}{n_{1}^{2}}K_{2}(2amn_{1}) +
\frac{(eB)^{2}}{24\pi^2} \sum_{n_{1}=1}^{\infty}
K_{0}(2amn_{1})+\nonumber\\ &-&\frac{7(eBa)^{4}}{1440\pi^2(am)^2}
\sum_{n_{1}=1}^{\infty}n_{1}^{2}K_{2}(2amn_{1})-\frac{2m^{4}}{\pi^2}
\sum_{n_{1},n_{2}=1}^{\infty} \frac{K_{2}(\sqrt{(2amn_{1})^{2} +
(\beta mn_{2})^{2}})}{(2amn_{1})^{2}+(\beta
mn_{2})^{2}}+\nonumber\\
&-&\frac{7(eBa)^{4}}{2880\pi^2(am)^{4}}\sum_{n_{1},n_{2}=1}^{\infty}
[(2amn_{1})^{2}+(\beta mn_{2})^{2}]K_{2}(\sqrt{(2amn_{1})^{2}+
(\beta mn_{2})^{2}})
\label{elfra}
\end{eqnarray}
and the pressure is given by:
\begin{eqnarray}
p_C(a,B,\beta)&=& \frac{180}{\pi^4}(am)^{2} \sum_{n_{1}=1}^{\infty}
\frac{1}{n_{1}^{2}}\biggl [
K_{2}(2amn_{1})+\frac{1}{3}(2amn_{1})K_{1}(2amn_{1})\biggl ]+\nonumber\\
&+&\frac{10(eBa^2)^{2}}{\pi^4}\sum_{n_{1}=1}^{\infty} \biggl [
K_{0}(2amn_{1}) - (2amn_{1})K_{1}(2amn_{1}) \biggl ] + \nonumber\\
&-& \frac{7}{6}\frac{(eBa^2)^{4}}{{\pi}^{4}(am)^{2}}\sum_{n_{1}=1}^{\infty}
n_{1}^{2}\biggl [ K_{2}(2amn_{1})-(2amn_{1})K_{1}(2amn_{1})\biggl
] +\nonumber\\
&-& \frac{480(am)^{4}}{\pi^{4}}\sum_{n_{1},n_{2}=1}^{\infty} \biggl
\{ \frac{K_{2}(\sqrt{(2amn_{1})^{2}+(\beta mn_{2})^{2}})}
{(2amn_{1})^{2}+(\beta mn_{2})^{2}}+\nonumber\\
&+& (2amn_{1})^{2} \biggl [ \frac{K_{1}
(\sqrt{(2amn_{1})^{2}+(\beta mn_{2})^{2}})}
{[(2amn_{1})^{2}+(\beta mn_{2})^{2}]^{3/2}} +
4\frac{K_{2}(\sqrt{(2amn_{1})^{2}+(\beta mn_{2})^{2}})}
{[(2amn_{1})^{2}+(\beta mn_{2})^{2}]^2}\biggl ] \biggl
\}+\nonumber\\
&+&\frac{20(eBa^2)^2}{\pi^4}\sum_{n_{1},n_{2}=1}^{\infty}\biggl [
K_{0}(\sqrt{(2amn_{1})^{2}+(\beta mn_{2})^{2}})+\nonumber\\
&-&(2amn_{1})^{2}\frac{K_{1}(\sqrt{(2amn_{1})^{2}+(\beta mn_{2})^{2}})}
{\sqrt{(2amn_{1})^{2}+(\beta mn_{2})^{2}}}\biggl ]+\nonumber\\
&-& \frac{7(eBa^{2})^4}{12\pi^4
(am)^4}\sum_{n_{1},n_{2}=1}^{\infty} \biggl \{[(2amn_{1})^{2}+(\beta
mn_{2})^{2}]K_{2}(\sqrt{(2amn_{1})^{2}+(\beta
mn_{2})^{2}})+\nonumber\\
&+& (2amn_{1})^{2} \biggl [\sqrt{(2amn_{1})^{2}+(\beta mn_{2})^{2}}
K_{1}(\sqrt{(2amn_{1})^{2}+(\beta mn_{2})^{2}})+\nonumber\\
&+&4K_{2}(\sqrt{(2amn_{1})^{2}+(\beta mn_{2})^{2}})\biggl ] \biggl \}
\end{eqnarray}
In this expression we can see the contribution of a weak external
field on the Casimir pressure as well as the combined effect of
the weak field and the finite temperature.
\subsection*{Acknowledgments}
M. V. C.-P. and C. F. would like to acknowledge CNPq (The National
Research Council of Brazil) for partial financial support. M. R. N.
acknowledges CAPES for partial financial support.
\vfill\eject

\end{document}